\documentclass[%
 reprint,
 amsmath,amssymb,
 aps,
]{revtex4-2}

\usepackage{graphicx}
\usepackage{dcolumn}
\usepackage{bm}
\usepackage{overpic}
\DeclareMathOperator{\ODESolve}{ODESolve}

\begin{document}

\preprint{APS/123-QED}

\title{Data-Driven Surrogates of Rotating Detonation Engine Physics with Neural ODEs and High-Speed Camera Footage}

\author{James Koch}
 \email{jvkoch0@gmail.com}

\date{\today}

\begin{abstract}
Interacting multi-scale physics present in the Rotating Detonation Engine lead to diverse nonlinear dynamical behavior, including combustion wave mode-locking, modulation, and bifurcations. In this work, surrogate models of the RDE physics, including combustion, injection, and mixing, are sought that can reproduce the observed behavior through their interactions. These surrogate models are constructed and trained within the context of Neural ODEs evolving through the latent representation of the waves: the traveling wave coordinate $\xi = x - ct + a$. Shown is that the multi-scale nature of the physics can be successfully separated and analyzed separately, providing valuable insight into the fundamental physical processes of the RDE. 
\end{abstract}

\maketitle


\section{Introduction} \label{sec:intro}
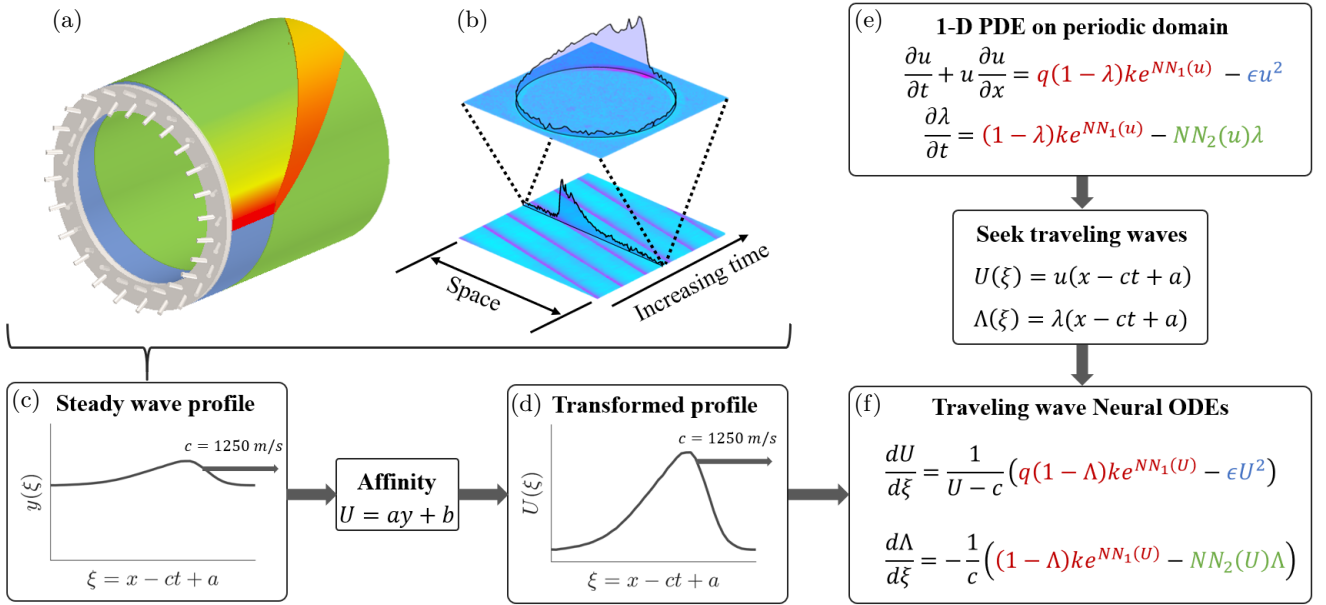
\begin{figure*}[]
        \centering
        \begin{overpic}[width=1.0\linewidth]{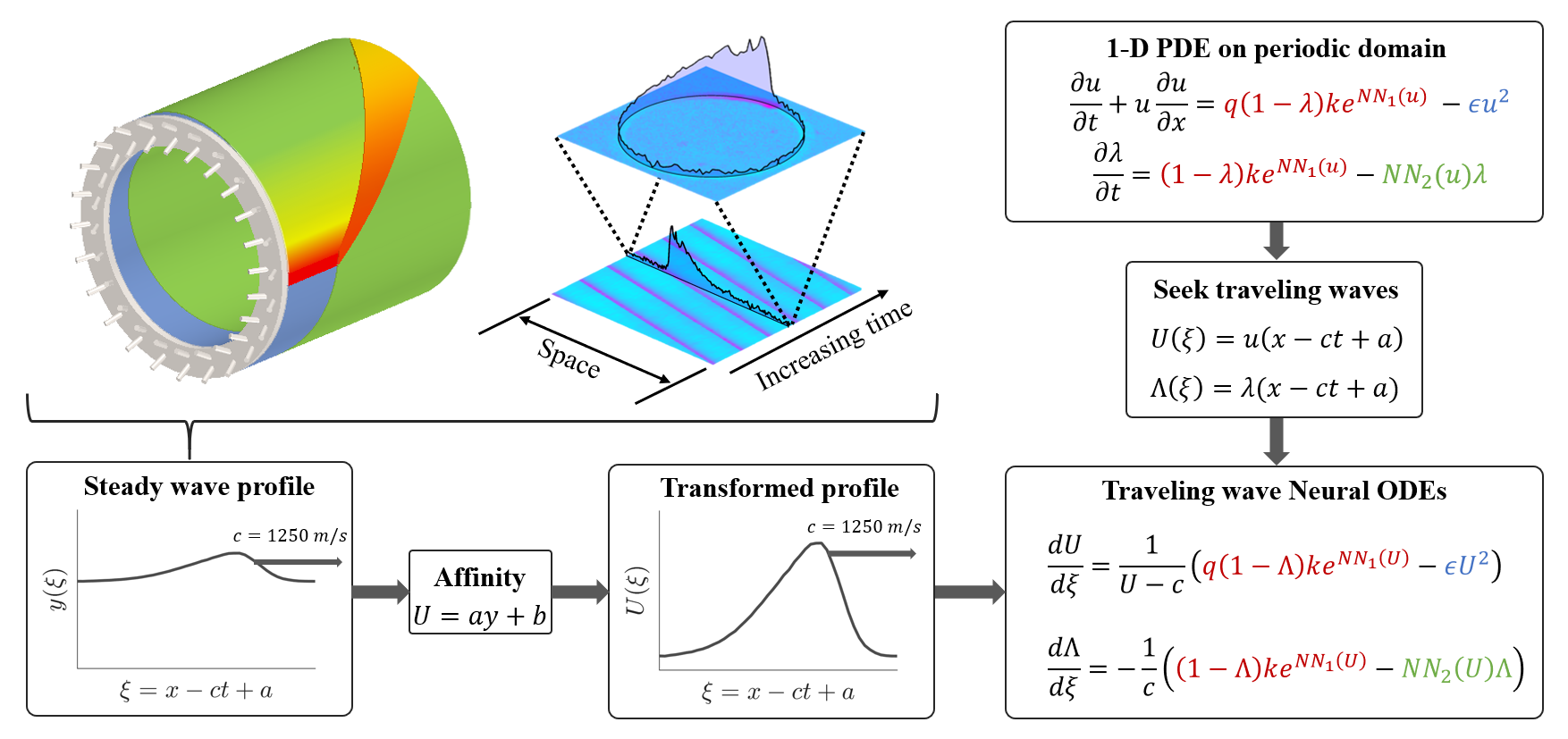}  
        \put(5,44){(a)}
        \put(35,44){(b)}
        \put(2,15.8){(c)}
        \put(39,15.5){(d)}
        \put(64.5,44){(e)}
        \put(64.5,15.5){(f)}

	    \end{overpic}  
	    \caption{A notional Rotating Detonation Engine (RDE) is shown in (a). Fuel and oxidizer enter injection ports on the head-end of the device and mix to form a combustible mixture. A number of detonation waves traverse the periodic combustion chamber consuming the newly injected and mixed propellant. The high-enthalpy exhaust flow is expelled rearwards. A common method for observing the detonation wave kinematic behavior is high-speed camera imaging (visible spectrum). High-speed camera footage provides a low-noise and well-resolved space-time history of the traveling waves (b). Snapshots of the annular combustion chamber (128-square pixel images) can be converted into 1-D vectors containing the pixel-integrated luminosity about the annulus. These snapshots can be stacked to form a space-time diagram of the experiment. For steady-state operation, multiple camera frames can be phase-averaged to produce a representative steady wave luminosity profile (c). In (e), a wave equation describing the damped-driven activator-inhibitor behavior of the RDE is written for a 1-D periodic domain. Trained neural networks are sought to approximate single-step Arrhenius kinetics (shown in red) and the mixing sub-model (green) subject to dissipation (blue). In seeking traveling wave solutions, the partial differential equation is transformed to a system of ordinary differential equations (ODEs) evolving through the coordinate $\xi=x-ct+a$ (f). This model is fit to an affine transformation of the high-speed camera footage (d) through auto-differentiation.}
		\label{fig:intro}
\end{figure*}

The \textit{Rotating Detonation Engine} (RDE) is a novel propulsion or stationary power-producing device that seeks to exploit the advantageous thermodynamic properties of constant-volume combustion through continuously rotating detonation waves \cite{Nordeen_2014}. RDE combustors feature periodic combustion chambers (e.g. an annular combustion chamber, such as the idealized combustor pictured in Fig. \ref{fig:intro}a) with fuel and oxidizer injected at the head-end of the device. One or more traveling combustion waves consume the mixed fuel and oxidizer mixture and expel the exhaust rearwards, producing thrust or otherwise high-enthalpy flow from which work can be extracted. The traveling combustion waves are the result of the highly dissipative, multi-scale, and nonlinear balance physics associated with the RDE: energy gained through combustion competes with dissipation through exhaust, while the fast combustion process is balanced by (and subject to) the slow physics of injection and mixing \cite{Koch_2020,Koch2020c,Koch2020b}. 

These multi-scale physics pose a significant challenge to the engineering task of RDE design, operation, and control. These tasks typically are rooted in high-fidelity computational fluid dynamic simulations with detailed combustor geometry, chemical kinetics, and turbulence models that can properly resolve all associated physical scales present in the RDE (see \cite{Cocks_2016}, for example). Because the physics span several orders of magnitude (from sub-microsecond to milliseconds), these simulations are computationally expensive or intractable. Furthermore, validation of these simulations is difficult because of the lack of well-resolved, low-noise, and relevant data from experiments due to the harsh sensing environment. However, high-speed camera footage  of RDEs (such as the data in Fig. \ref{fig:intro}b) provides a readily-obtained and spatially-resolved space-time history of the kinematics of the combustion waves. Chemiluminescence as observed by high-speed cameras does not have a direct relationship to engineering quantities of interest (pressure or temperature, for example), although many have noted basic qualitative properties of this observable that may correlate with fluid properties \cite{Bennewitz_2019,Journell2020,Huang2021}. For example, greater luminosity generally implies higher temperature, pressure, and reaction rate. 

In this work, chemiluminescence measurements obtained from high-speed camera footage is used to construct \textit{surrogate models} of the multi-scale nonlinear phenomena that contribute to the characteristic traveling wave behavior of RDEs within the context of a specific RDE experiment. These end-use goal of these surrogate models is to provide estimates of physical scales and inform how the interactions of these scales result in the observed behavior. This work is critically enabled by: (i)  the observation that steadily propagating rotating detonation waves can be completely described by a system of Ordinary Differential Equations (ODEs) evolving in the traveling wave coordinate $\xi = x - ct + a$, where $c$ is the speed of the assumed right-running wave and $a$ is an arbitrary spatial offset \cite{Koch_2021}, (ii) the observation that pixel-integrated luminosity is linear with speed (Section \ref{sec:transition}), and (iii) \textit{Neural ODEs} (NODEs) and their generalization; \textit{Univeral ODEs} \cite{Rackauckas2020}. A model \textit{activator-inhibitor} dynamical system (Fig. \ref{fig:intro}e) evolving through the traveling wave coordinate $\xi$ is constructed with basic assumptions regarding the essential RDE balance physics (Fig. \ref{fig:intro}f). Assuming that luminosity is a sufficient surrogate measurement for the \textit{activator} (Fig. \ref{fig:intro}c-Fig. \ref{fig:intro}d), this dynamical system is fit to experimentally-obtained high-speed camera footage.

In Section \ref{sec:experiments}, the experimental apparatus and data are presented. In Section \ref{sec:analysis}, the model form is constructed and the computational framework is prescribed. The results and interpretation of the fit model are presented in Section \ref{sec:results} followed by discussion and conclusions in Section \ref{sec:discussion}.

\section{Experiments} \label{sec:experiments}

High-speed camera footage from two experiments is presented in this section: (i) the transition of two co-rotating to a single wave (Section \ref{sec:transition}), and (ii) four steadily propagating combustion waves (Section \ref{sec:modeLocked}). The former experiment is used to establish speed-amplitude properties of as-observed luminosity. The latter experiment is the source of data for model training in Section \ref{sec:analysis}. Both experiments were performed at the University of Washington High Enthalpy Flow Laboratory with a 76-mm flowpath outside diameter RDE with a mixture of gaseous methane and oxygen. The engine test stand is a closed vacuum chamber with a large exhaust plenum and in-line optical access for direct line-of-sight to the exit of the combustor. A complete description of the test stand experimental apparatus is given in \cite{Koch_2019} and \cite{Koch_2019a}, respectively. Each experiment was imaged with a Phantom Camera v2511 high speed camera at 240,000 frames per second at a resolution of 128x128 pixels. As shown in Fig. \ref{fig:intro}b,  the high-speed camera footage from experiments can be manipulated to form space-time diagrams of the wave kinetics. These histories are the primary source of data for this work. 

\subsection{Speed-Luminosity Relationship} \label{sec:transition}

\begin{figure}[]
        \centering
        \begin{overpic}[width=1.0\linewidth]{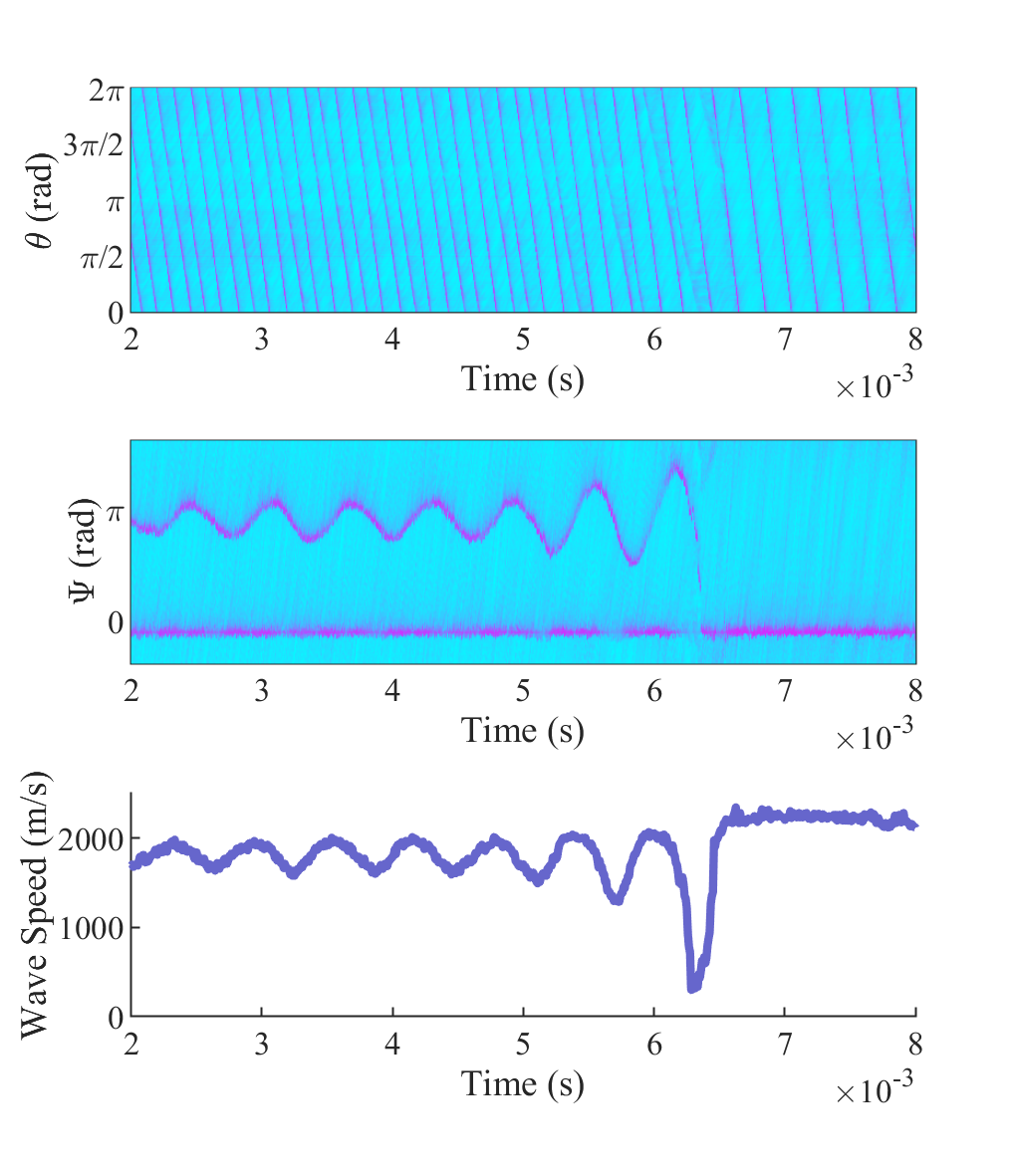}   
	    \end{overpic}  
	    \caption{An experimental space-time history of a destructive wave bifurcation is shown in (a) in the laboratory reference frame. Shifting the data in (a) into the wave-attached reference frame yields (b) where the evolution of the phase difference between the waves is clearly observed. In this experiment, the two waves oscillate with greater amplitude until one wave overruns the other. In (c), the speed of the tracked wave in (b) is shown.}
		\label{fig:transition}
\end{figure}

Wave unsteadiness is one of the hallmark features of the RDE. This unsteadiness is manifested as wave modulation, bifurcations to more or fewer waves, and/or chaotic propagation. Figure \ref{fig:transition} depicts the space-time history of an experiment during mode transition from two co-rotating waves to a single traveling wave. In Fig. \ref{fig:transition}a, the post-processed high-speed camera footage is shown in the laboratory reference frame. The spatial dimension is normalized to reflect the $2\pi$ periodicity of the combustion chamber. In Fig. \ref{fig:transition}b, the data in Fig. \ref{fig:transition}a is shifted into the wave-attached reference frame. In this frame, the phase difference between the two waves is highlighted. This phase difference oscillates about $\pi$ radians with growing amplitude until one wave overtakes the other. Such mode transitions are common during the ramp-down of propellant feed at the end of each experiment: each wave competes for the increasingly scarce injected propellant until only one wave can be supported by the reduced inflow.

\begin{figure}[]
        \centering
        \begin{overpic}[width=1.0\linewidth]{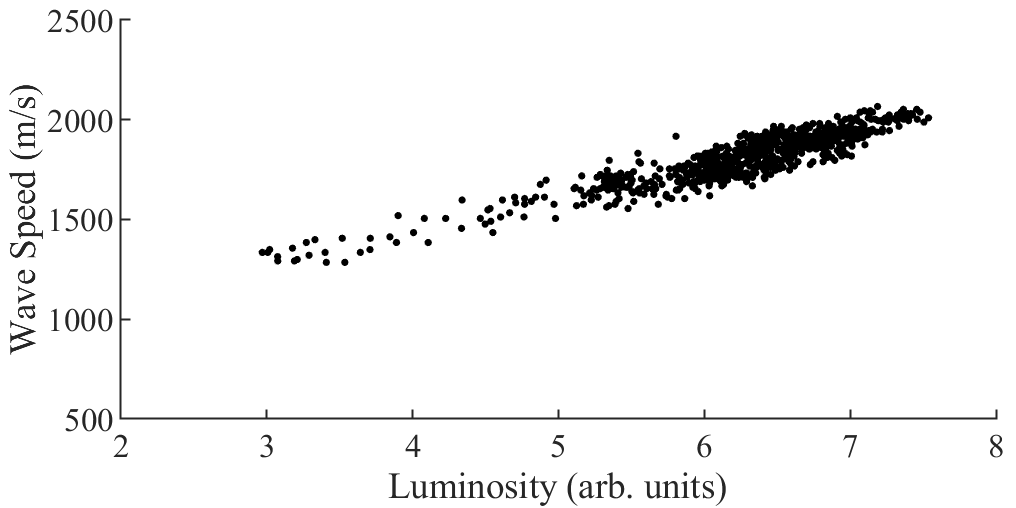}   
	    \end{overpic}  
	    \caption{The relationship between instantaneous wave speed and camera-observed luminosity is linear. Data points correspond to the wave amplitude along the line $\Phi=0$ in Fig. \ref{fig:transition}b and the corresponding instantaneous wave speed in Fig. \ref{fig:transition}c.}
		\label{fig:linearRelationship}
\end{figure}

The oscillations in phase difference of the waves are also accompanied with oscillations in wave speed and observed luminosity. Figure \ref{fig:transition}c shows the evolution of speed through the modal transition. The relationship between camera-observed luminosity and wave speed is approximately \textit{linear}, as shown in Fig. \ref{fig:linearRelationship}. Linear speed-amplitude relationships are pervasive in nonlinear wave physics; examples include the Korteweg-de Vries Equation \cite{Miura1976} and Burgers' Equation \cite{Su1969} with each possessing the nonlinear advection term $u \frac{\partial u}{\partial x}$. Thus, the amplitude of this observable is dependent on the speed of the waves and this property should be retained in model construction.

\subsection{Steady Mode-locked Waves} \label{sec:modeLocked}

\begin{figure}[]
        \centering
        \begin{overpic}[width=1.0\linewidth]{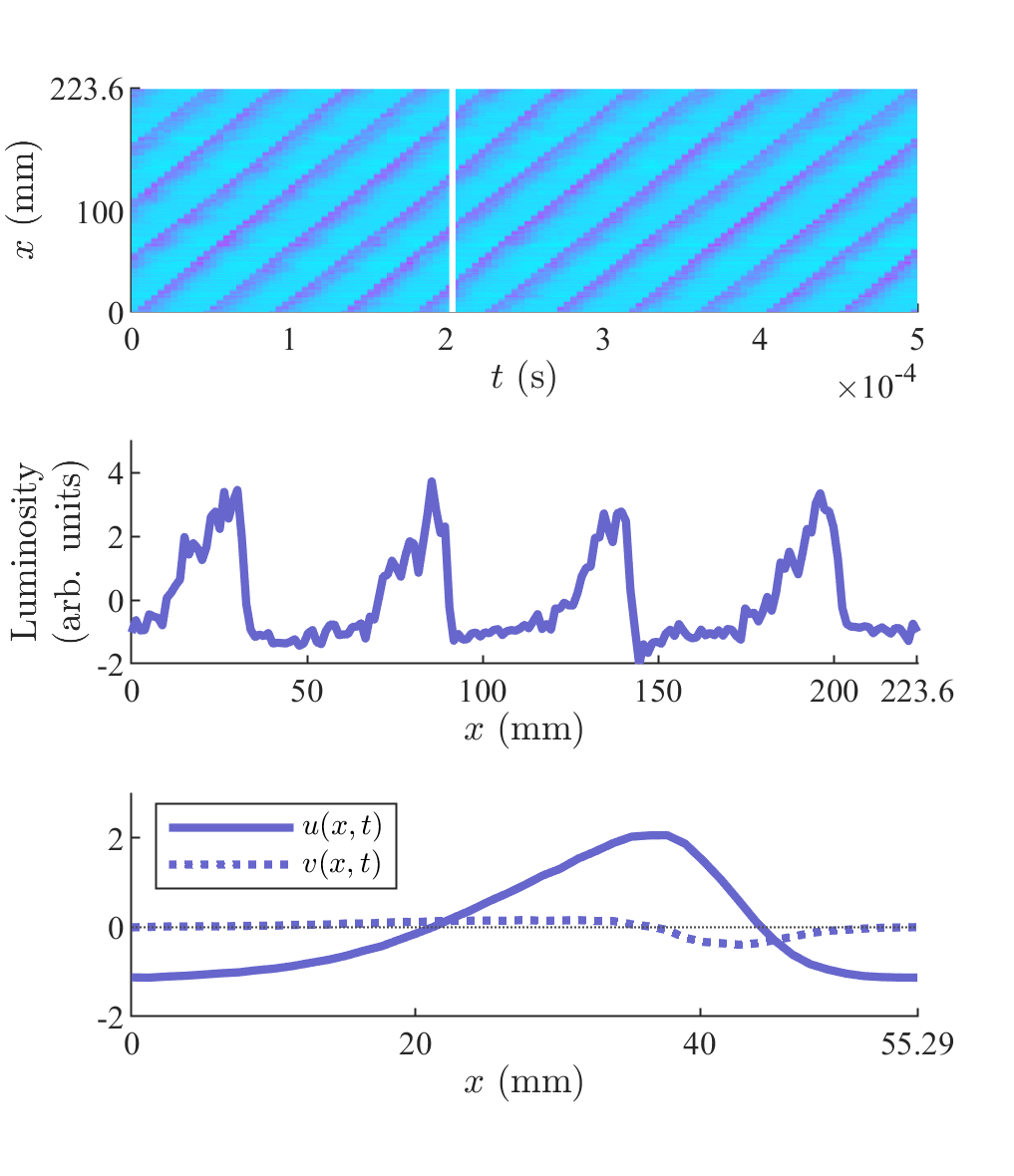}   
	    \end{overpic}  
	    \caption{In (a), displayed are four mode-locked rotating detonation waves as observed by a high-speed camera. In (b), an instantaneous snapshot of the system at $t=2\cdot 10^{-4}$ seconds is shown exhibiting the characteristic RDE 'sawtooth' wave profile. Because these waves are approximately steady, they can be phase-averaged to produce a low-noise representative profile, such as that which is shown in (c).}
		\label{fig:modeLockedData}
\end{figure}

Figure \ref{fig:modeLockedData} shows high-speed camera footage from an experiment with four steadily propagating combustion waves (global equivalence ratio of 0.32 and mass flow rate of 226 g/s). In (a), the space-time history is shown in the laboratory reference frame. The spatial coordinate bounds correspond to the length of the annular chamber. In (b), an instantaneous circumferential 1-D snapshot of the luminosity wave profile (corresponding to the vertical white bar in (a)) is shown. The slopes of the stripes in (a) correspond to the wave speed for this portion of the experiment. For this case, the measured velocity is 1250 m/s. For the given methane-oxygen input mixture, this corresponds to 65\% of the theoretical detonation speed at standard temperature and pressure.

Because the waves are approximately steady, they can be phase-averaged to produce a representative wave profile. This profile is given in Fig. \ref{fig:modeLockedData}c (solid line) along with its spatial derivative (dashed line) for one period. This reduced-noise luminosity profile is the training data for Sections \ref{sec:analysis} and \ref{sec:results}.

\section{Methodology} \label{sec:analysis}

\subsection{Physics-Informed Model Architecture}
The RDE is a damped-driven activator-inhibitor system. The propagating combustion wave fronts are \textit{autosolitons} that result from the \textit{local} balance of nonlinearity (self-steepening of waves), gain (combustion), and dissipation (exhaust) subject to the slow-scale gain recovery (injection and mixing). The \textit{Rotating Detonation Analog} \cite{Koch_2020} relates mathematical approximations of these physical processes and their associated scales in a compact framework - a two-component 1-D partial differential equation - that is amenable to standard analysis techniques:

\begin{equation}
    \frac{\partial u}{\partial t} + u \frac{\partial u}{\partial x} = q(1-\lambda) \omega(u) -\epsilon u^2
\end{equation}
\begin{equation}
    \frac{\partial \lambda}{\partial t} = (1-\lambda)\omega(u) - \beta(u)\lambda ,
\end{equation}
where $u(x,t)$ is a spatially and temporally-variable analogous to specific internal energy, $\lambda(x,t)$ is a combustion progress variable, $q$ is heat release, $\omega(u)$ is the chemical kinetic sub-model, $\epsilon$ is the loss coefficient, and $\beta(u)$ is the injection and mixing sub-model. In general, $\omega(u)$ and $\beta(u)$ are parameterized functions of $u(x,t)$.

This analog system is able to qualitatively reproduce the nonlinear dynamics observed in experiments, including wave mode-locking, modulation, and bifurcations. Additionally, this analog model contains the nonlinear advection term $u \frac{\partial u}{\partial x}$ responsible for linear speed-amplitude behavior and shock formation. Thus, expected is that a generalized model of this form can be fit to an affine transformation of the phase-averaged experimental luminosity observations presented in Fig. \ref{fig:modeLockedData}c. However, the exact functional forms of these physics sub-models may be incorrect or otherwise improperly specified. Therefore, instead the following system is fit to the experimental luminosity wave profile:

\begin{equation} \label{eq:u}
    \frac{\partial u}{\partial t} + u \frac{\partial u}{\partial x} = q(1-\lambda) k\exp(NN_1(u)) -\epsilon u^2
\end{equation}
\begin{equation} \label{eq:lambda}
    \frac{\partial \lambda}{\partial t} = (1-\lambda)k\exp(NN_1(u)) - NN_2(u)\lambda ,
\end{equation}
where $k\exp(NN_1)$ is a single-step Arrhenius surrogate model for chemical kinetics with a Neural Network (NN) and $NN_2$ is the Neural Network surrogate model for injection and mixing. The Burgers' flux and dissipation terms are unchanged as these functional forms are derived as opposed to modeled. 

Equations \ref{eq:u} and \ref{eq:lambda} are transformed into Ordinary Differential Equations (ODEs) by seeking traveling wave solutions with the transformation into the coordinate $\xi = x - ct + a$. The model form becomes:

\begin{equation} \label{eq:UU}
\begin{split}
    \frac{dU}{d\xi} = \frac{1}{U-c}\left( q(1-\Lambda)k\exp(NN_1(U)) -\epsilon U^2 \right) \\ = g(U,\Lambda;\theta)
\end{split}
\end{equation}
\begin{equation} \label{eq:LLambda}
\begin{split}
   \frac{d\Lambda}{d\xi} = -\frac{1}{c}\left( (1-\Lambda)k\exp(NN_1(U)) - NN_2(U)\Lambda \right) \\ = h(U,\Lambda;\theta),
\end{split}
\end{equation}
with $u(x,t) = U(\xi=x-ct+a)$ and $\lambda(x,t) = \Lambda(\xi=x-ct+a)$. The model equations \ref{eq:UU} and \ref{eq:LLambda} are now amenable to standard machine learning techniques for dynamical systems. However, instead of evolving through time, here the model evolves through the time-like coordinate $\xi$. 

For consistency, the representative phase-averaged wave profile (Fig. \ref{fig:modeLockedData}c) needs to be transformed into $\xi$. This transformation is done implicitly by setting the spatial offset $a$ to be exactly equivalent to $ct$; thus $\xi = x - ct + a = x$. Note that this implicit transformation only holds for individual snapshots of the system, but that in the case for steadily propagating waves, each snapshot is identical up to an arbitrary spatial shift.  

\subsection{Regression with Neural ODEs}
Neural ODEs evolve a state $\mathbf{z}$ through an independent variable ($\xi$ in this case) through a prescribed input-output mapping defined by a neural network or composite function containing a neural network, $f$:

\begin{equation} \label{eq:node1}
    \mathbf{z}(\xi_{end}) = \mathbf{z}(\xi_{0}) + \int_{\mathbf{z}(\xi_{0})}^{\mathbf{z}(\xi_{end})} f(\mathbf{z}(\xi);\theta) d\xi ,
\end{equation}
where $\theta$ are the trainable parameters of the model. Equation \ref{eq:node1} - an initial value problem - can be solved with standard numerical ODE solvers:

\begin{equation}
    \mathbf{z}(\xi_{end}) = \ODESolve(f,\mathbf{z_0},\xi_0,\xi_{end};\theta) .
\end{equation}
The model parameters $\theta$ are optimized by backpropagating residuals through the numerical ODE solver to provide necessary gradients.

The loss function to be minimized compares the phase-averaged luminosity profile of Fig. \ref{fig:modeLockedData}c to numerically integrated neural ODE waveforms. Note that because the only observable is luminosity, or the activator, there is no constraint on $\Lambda$ dynamics. Furthermore, the necessary translation and scaling of the luminosity data to fit the physical and mathematical assumptions (satisfying Burgers' flux, for example) is unknown. Therefore, in addition to optimizing the model parameters, an affine transformation of the data must also be learned. This is accomplished by encoding a proportionality constant $m$ and offset $n$ into the loss function:

\begin{equation} \label{eq:odesolve}
\begin{split}
    \begin{bmatrix}
    U(\xi) \\
    \Lambda(\xi)
    \end{bmatrix} = \ODESolve \left(\begin{bmatrix} g \\ h \end{bmatrix},\begin{bmatrix} my_0 + n \\ \Lambda_0 \end{bmatrix}, \xi_0, \xi_{end};\theta \right) ,
\end{split}
\end{equation}

\begin{equation}
    \mathcal{L} = \sum_j \left( y(\xi_j) - \frac{U(\xi_j)-n}{m} \right) ,
\end{equation}
where $y(\xi)$ is the experimental data, $\xi_0 = 0$ meters, $\xi_{end} = 0.0559$ meters (length of one period), $\Lambda_0$ is a user-specified hyper-parameter setting a reference value for the combustion progress variable ($\Lambda = 0.9$ in this work), and the left-hand side of Eq. \ref{eq:odesolve} represents output data streams for $U$ and $\Lambda$ on equally-spaced evaluation points $j$ corresponding to the training data. 

The model was constructed and trained in the Julia computing ecosystem \cite{Rackauckas2019}. The ODE solver used was a 5th-order Runge-Kutta integrator. The neural networks used each consist of a single sigmoid-activated layer of dimension 3 and a linear output layer. The BFGS optimization routine was used to select parameters subject to the loss function. For this exploratory study, convergence was deemed sufficient when the error plateaued. 

\section{Results} \label{sec:results}

The regression is performed with the knowledge of (i) the wave speed and (ii) the fractional Chapman-Jouguet detonation velocity. For this experiment, the waves traveled at 1250 m/s corresponding to 65\% of the theoretical detonation velocity. The RDE analog has a corresponding detonation velocity equal to twice the heat release; $D_{CJ} = 2q$ \cite{Majda1981}. The heat release parameter is set to $q = 932$ such that $D_{CJ} = 1864$ m/s, matching the detonation velocity of the propellant mixture in the experiment for this fueling condition. The unknown parameters are $a$,$b$,$\epsilon$ and $k$ in addition to the neural network weights. After multiple successful training trials and network architecture permutations, it was found that the neural network approximating the injection and mixing processes defaults to a constant output value without compromising model accuracy. This network has therefore been replaced with a single trainable parameter, $s$, for these results. 

\begin{figure}[]
        \centering
        \begin{overpic}[width=1.0\linewidth]{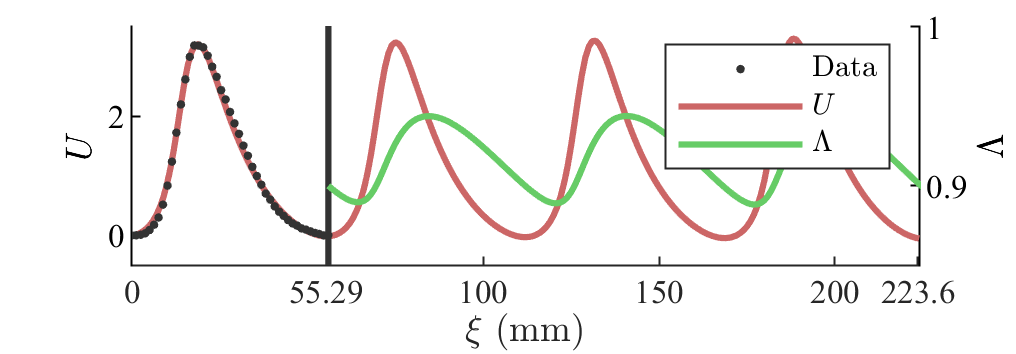}   
	    \end{overpic}  
	    \caption{Experimental luminosity observations are shown to the left of the vertical black bar at 55.29 mm. The reconstruction of the data is given by the under-laid red trace with extrapolation to the right of the black bar. The model output additionally provides an estimate for the combustion progress variable dynamics as shown in green. The neural ODE model is able to successfully reproduce the expected limit cycle behavior.}
		\label{fig:neuralODE}
\end{figure}

Figure \ref{fig:neuralODE} shows the reconstruction and continuation of the experimental data with the trained nerual ODE model. The training data corresponds to the discrete plot markers to the left of the vertical black bar. The solid red trace is the neural ODE reconstruction of the data. The green trace corresponds to the combustion progress variable dynamics. The model is successful in reproducing the limit cycle behavior of this experiment, including the unobservable state variable $\Lambda$. 

\begin{figure}[]
        \centering
        \begin{overpic}[width=1.0\linewidth]{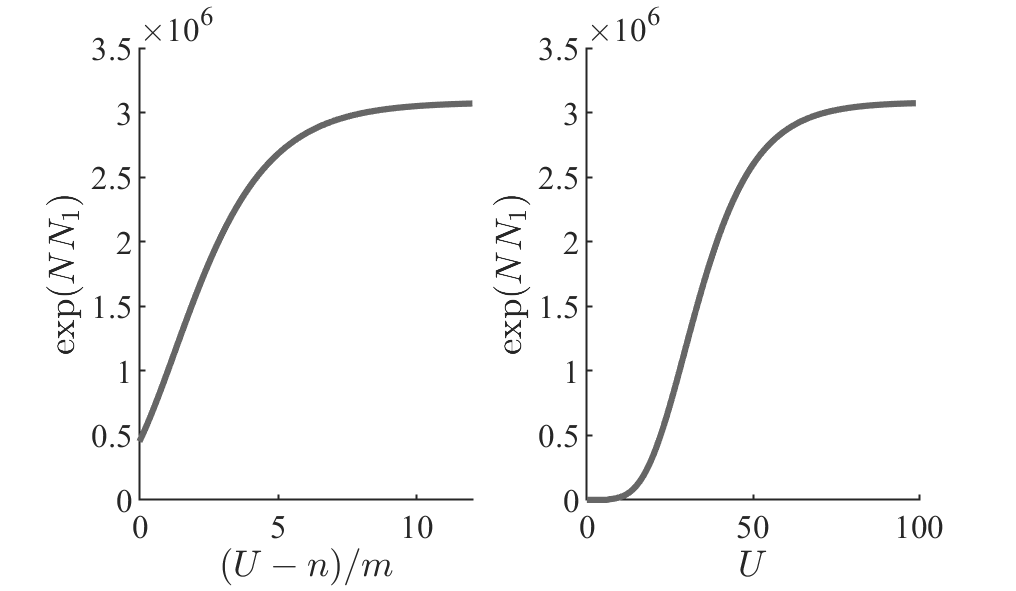}  
        \put(3,56){(a)}
        \put(48,56){(b)}
	    \end{overpic}  
	    \caption{Response of the surrogate chemical kinetic model to luminosity - the assumed activator. In (a), the abscissa is scaled to match the original data. In (b), the abscissa is displayed after the scaling ($m$) and translation ($n$).}
		\label{fig:kinetics}
\end{figure}

In addition to reproducing the data, the model can be further dissected to give physical insights about this particular experiment. Figure \ref{fig:kinetics} shows the trained surrogate model for chemical kinetics \textit{based on activation through luminosity}. As expected for an assumed single-step chemical reaction model, the reaction rate increases with luminosity, eventually plateauing in the limit of large luminosity. These reaction rates can be directly compared with another model output; the propellant regeneration rate constant $\tau = 1/s = 1.67 \cdot 10^{-4}$s. Thus, the model is able to successfully separate the wave physics into its constitutive slow (injection and mixing) and fast (combustion) physics.

\begin{table}[t]
\begin{tabular}{ccccc}

 $k$ & $\epsilon$  & $s$  & $m$ & $n$ \\
 \hline \hline
 0.0687 & 5920 & 5980 & 6.16 & 21.6 \\
 \hline

\end{tabular}
\caption{Trained model parameters.}
\label{tab:my-table}
\end{table}

\section{Conclusion} \label{sec:discussion}

The RDE possess a cascade of interacting physical scales that complicate the physics investigation and engineering tasks associated with these engines. In this work, the traveling wave structure of the steady-state behavior of the RDE is exploited to construct and train an interpretable, physics-informed neural ODE system. This model evolves through the traveling wave coordinate $\xi = x-ct+a$ as opposed to coupled space-time dynamics. The resulting model is capable of successfully separating the slow and fast physics directly from luminosity measurements as recorded by a standard laboratory-grade high-speed camera. The implications are threefold: (i) the ability to disentangle the physical scales of a particular experiment and/or engine enables the forward analysis tasks of wave existence, stability, and the like, (ii) the estimates of physical scales can be tracked and associated with a specific test article, providing a basis for a Rotating Detonation Engine \textit{Digital Twin} and in situ non-intrusive diagnostics, and (iii) the surrogate models can be re-introduced to the full partial differential equation and used to investigate operability, especially if parameterized by operating conditions. 

\begin{acknowledgments}
All experiments were performed at the University of Washington High Enthalpy Flow Laboratory in Seattle, Washington.
\end{acknowledgments}

\bibliography{rde}

\end{document}